# Two-color ionization of hydrogen by short intense pulses

S.Bivona, G. Bonanno, R.Burlon and C. Leone

CNISM and Dipartimento di Fisica e Tecnologie Relative, Università degli Studi di Palermo, Viale delle Scienze Ed 18, 90128 Palermo Italy

Photoelectron energy spectra resulting by the interaction of hydrogen with two short pulses having carrier frequencies, respectively, in the range of the infrared and XUV regions have been calculated. The effects of the pulse duration and timing of the X-ray pulse on the photoelectron energy spectra are discussed. Analysis of the spectra obtained for very long pulses show that certain features may be explained in terms of quantum interferences in the time domain. It is found that, depending on the duration of the X-ray pulse, ripples in the energy spectra separated by the infrared photon energy may appear. Moreover, the temporal shape of the low frequency radiation field may be inferred by the breadth of the photoelectron energy spectra.

Corresponding author

Saverio Bivona
Dipartimento di Fisica e Tecnologie Relative, Viale delle Scienze – Ed. 18, I-90128 Palermo Italy
Phone: ++39-091-6615104
Fax: ++39-091-6615063
E-mail: bivona@unipa.it

1 Introduction

Ionization of atoms by soft x-ray radiation in the presence of a low-frequency intense radiation field has been the subject of both theoretical and experimental studies since two decades [1]. Calculations of energy spectra for the case of atomic systems irradiated by a low-intense x-ray field in the presence of an intense long infrared pulse described as a monochromatic field show photoelectron energy peaks evenly separated by the infrared photon energy [2,3]. Observations of this kind of sidebands in experiments of Helium ionization by simultaneous soft x-ray and infrared radiation have been reported on by Glover et al. in ref. [4] where the ponderomotive threshold shift due to the quiver energy provided by the infrared pulse to the photoelectron was measured too. Two color experiment have received a new impetus by the use of the free electron laser in Hamburg where ionization of helium atoms by extreme XUV pulse of tens of femtoseconds in the presence of intense synchronized picosecond optical laser have been studied [5,6].

Recent developments in laser technology have made it possible to produce short, high-power laser pulse with duration of few optical cycles, that have become available as research tools [7]. For not too short pulses, the electric field may be represented as a product of a monochromatic carrier wave and a positive-definite envelope function. One of the parameters characterizing this type of pulses is the so called carrier-envelope relative phase. By varying this parameter the temporal shape of the pulse may vary significantly, allowing coherent control and study of elementary atomic processes. An instance of application of this sources to the study of quantum fundamental processes has been recently given in attosecond double-slit experiments in the time-energy domain [8]. In these experiments, due to the highly nonlinear processes, the ionization occurs in time windows having the duration of attoseconds. By changing the relative carrier-envelope phase, the temporal shape of the field may be altered in such a way that the time windows may be ``open'' or ``closed'', controlling the recorded photoelectron spectra modulations that can be described in terms of quantum interference. Another application aimed at the study of fundamental processes is the ionization of atoms caused by subfemtosecond XUV pulses in the presence of a more intense femtosecond infrared radiation [9]. The burst of photoelectrons created by the x-ray pulse may be accelerated or decelerated by the femtosecond laser field, depending on the relative phase between the two radiation fields, as attosecond streak camera experiments have shown [10]. By observing the photoelectron energy spectra as a function of the delay between the x-ray and the assisting laser field, information on either the duration of the shortest pulse or the temporal shape of the electric field of the largest pulse may be obtained [11].

In the case of two color ionization, when the duration of the x-ray pulse shortens, a limit may be reached beyond which it is no longer possible to resolve the photoelectron energy in quantities comparable with the low-frequency photon energy. Therefore significant changes of the photoelectron energy spectrum are expected when the x-ray pulse duration is reduced to a temporal interval that is comparable, or even shorter, than the infrared field period [12] .

It is the aim of this paper to analyze and discuss some of the properties of the energy spectra of the photoelectrons emitted upon exposing a hydrogen atom to the influence of two radiation pulses. The first of them is an intense pulse (LF field) in the infrared region encompassing some radiation cycles; the other is a weaker XUV pulse (HF field) whose duration is assumed to range from a laser period fraction to many laser periods. Below it will be shown how both the shape and the photoelectron energy spectra breadth depend on the XUV timing and pulse duration.

In section 2 the photodetachment transition probability will be derived by treating the XUV radiation at the first order of the time-dependent perturbation theory, and taking into account, though approximately, the effect of the intense laser field on the electron motion by the so called Volkov-Coulomb approximation. In section 3 the photoelectron energy spectra obtained by using the Volkov-Coulomb approximation will be analyzed and, for the case of short XUV pulses, containing several HF radiation cycles, compared with those found by numerical integration of the three dimensional time dependent Schroedinger equation.

## 2. THEORY

The Hamiltonian of the atom in the presence of the two radiation fields, in the length gauge, is

$$H(t)=H_0+W_H(t)+W_L(t)=H_L(t)+W_H(t) \qquad (1)$$

with $H_0$ the field free Hamiltonian of the Hydrogen atom and $W_j(t)=e\mathbf{E}_j(t)\cdot\mathbf{r}$ (j=H,L) where

$$\mathbf{E}_j(t)=1/2\ E_{0j}\ [\exp(i\omega_j t)+c.c.]\ \hat{\mathbf{z}} \qquad (2)$$

are the oscillating parallel, linearly polarized, electric fields, $\mathbf{r}$ is the electron coordinates, $\omega_H$ and $\omega_L$ denote, respectively, the frequencies of the HF and LF fields which are taken in the dipole approximation and $\hat{\mathbf{z}}$ is an unitary vector directed along the z axis.

Assuming at the instant t' the atom in an eigenstate $\left|\Phi_{i,0}(t')\right\rangle$ of the field-free hamiltonian $H_0$, at the time t the state vector of the atom will be given by (hereafter, atomic units will be used)

$$\left|\Psi(t)\right\rangle = U(t,t')\left|\Phi_{i,0}(t')\right\rangle \qquad (3)$$

with $\left|\Psi(t)\right\rangle$ and U(t,t') satisfying

$$H(t)\left|\Psi(t)\right\rangle = i\frac{d}{dt}\left|\Psi(t)\right\rangle \qquad (4)$$

$$\left(H_L(t)-i\frac{d}{dt}\right)U(t,t') = -W_H U(t,t') \qquad (5)$$

and U(t',t')=1. By expressing U(t,t') in terms of the time development operator $U_L(t,t')$ associated to the hamiltonian in the presence of the low frequency field as

$$U(t,t') = U_L(t,t') - i\int_{t'}^{t} U_L(t,\tau)W_H(\tau)U(\tau,t')d\tau \qquad (6)$$

with

$$\left(H_L(t)-i\frac{d}{dt}\right)U_L(t,t') = 0\,, \qquad (7)$$

we find for $\left|\Psi(t)\right\rangle$ the following integral equation

$$\left|\Psi(t)\right\rangle = \Phi_{i,L}(t) - i\int_{t'}^{t} U_L(t,\tau)W_H(\tau)\left|\Psi(\tau)\right\rangle d\tau \tag{8}$$

where $\left|\Phi_{i,L}(t)\right\rangle = U_L(t,t')\left|\Phi_{i,0}(t')\right\rangle$ is the state vector of the atom which would evolve from the initial state $\left|\Phi_{i,0}(t')\right\rangle$ under the action of the LF field only. Before proceeding further, it is convenient to specify the sequence of the physical operations which give rise to the time evolution of the atom.

It is assumed that at the instant t', when the atom is in its ground state, both the radiation fields are turned on. The time of switching on is so large that the field may be considered monochromatic and that the atom follows adiabatically the fields so that, when they reach their full intensity, the atom is still in its ground state. After a long interval time t-t', the fields are switched off adiabatically.

The probability amplitude that at the time t the atom goes into a state of the continuum $\left|\Phi_{f,0}(t)\right\rangle$ is given by

$$A_{if} = \left\langle\Phi_{f,0}(t)\middle|\Psi(t)\right\rangle = \left\langle\Phi_{f,L}(t')\middle|\Phi_{i,0}(t')\right\rangle - i\int_{t'}^{t}\left\langle\Phi_{f,L}(\tau)\middle|W_H(\tau)\middle|\Psi(\tau)\right\rangle d\tau \tag{9}$$

where

$$\left|\Phi_{f,L}(\tau)\right\rangle = U_L(\tau,t)\left|\Phi_{f,0}(t)\right\rangle = \left|\Phi_{f,0}(\tau)\right\rangle - i\int_{t}^{\tau} U_0(\tau,t'')W_L(t'')\left|\Phi_{f,L}(t'')\right\rangle dt'' \tag{10}$$

is the atomic state in the continuum in the presence of the LF field only that, for $t\rightarrow\tau$, goes into the state $\left|\Phi_{f,0}(\tau)\right\rangle$, and $U_0(t,\tau)=\exp[-iH_0(t-\tau)]$ is the field-free time-development operator. Upon substitution of Eq. (10) into Eq. (9), we find ($i\neq f$)

$$A_{if} = -i\int_{t'}^{t}\left\langle\Phi_{f,L}(\tau)\middle|W_L(\tau)\middle|\Phi_{i,0}(\tau)\right\rangle d\tau - i\int_{t'}^{t}\left\langle\Phi_{f,L}(\tau)\middle|W_H(\tau)\middle|\Psi(\tau)\right\rangle d\tau \tag{11}$$

This last expression of the amplitude transition contains information on all the channels which open when the atom is hit by the fields, and shows that there are different paths leading to the ionization process. In fact, the first term in Eq. (11) is associated to the direct multiphoton ionization of the atom by the intense LF field, while the other takes into account the simultaneous absorption and emission of both the LF and HF photons. By expanding $\left|\Psi(t)\right\rangle$ in powers of $W_H$, $A_{if}$ may be written as

$$A_{if} = -i\int_{t'}^{t}\left\langle\Phi_{f,L}(\tau)\middle|W_L(\tau)\middle|\Phi_{i,0}(\tau)\right\rangle d\tau - i\int_{t'}^{t}\left\langle\Phi_{f,L}(\tau)\middle|W_H(\tau)\middle|\Phi_{i,L}\right\rangle d\tau + O[(W_H)^K], \quad (K\geq 2) \tag{12}$$

Now we concentrate our attention on the ionization process due to the absorption of a single linearly polarized HF frequency photon accompanied by the exchange of any number of LF photons. Hence the first term of Eq. (12) will no longer be considered, and the second term will be taken in the rotating wave approximation writing $W_H=(1/2)\mathbf{E_{0H}}\cdot\mathbf{r}\exp(-i\omega_H t)$

As no analytical solutions of the Schroedinger equation of the atom in the presence of an electromagnetic field are known, we are faced with the problem of approximating the wavefunction $\Phi_{i,L}(\tau)$ and $\Phi_{f,L}(\tau)$ entering Eq. (12).

If it is assumed that during the time interval t-t' the initial atomic state is not appreciably depleted by the action of the intense laser field, the state $\left|\Phi_{i,L}(\tau)\right\rangle$ may be approximated as

$$\left|\Phi_{i,L}(\tau)\right\rangle \approx \left|\Phi_{i,0}(\tau)\right\rangle \tag{13}$$

and $A_{if}$ may be approximated by

$$A_{if} \approx -i\int_{t'}^{t} \left\langle \Phi_{f,L}(\tau)\right| W_H(\tau) \left|\Phi_{i,0}(\tau)\right\rangle d\tau \tag{14}$$

As far as the field dressed continuum states $\left|\Phi_{f,L}\right\rangle$ are concerned, as no analytical solution is until now known, they will be described by the so called Volkov-Coulomb wavefunction characterized by the canonical momentum **k**

$$\Phi_{f,L}(\mathbf{r},t) = \Phi^{-}_{\mathbf{k}+\mathbf{k}_L}(\mathbf{r})\exp\{i\mathbf{k}_L\cdot\mathbf{r}\}\exp\left\{-\frac{i}{2}\int^{t}[\mathbf{k}+\mathbf{k}_L(t)]^2\,dt'\right\} \tag{15}$$

where $\Phi^{-}_{\mathbf{q}}(\mathbf{r})$ is the incoming Coulomb wave describing an electron with asymptotic momentum **q**, $\mathbf{k_L}=(1/c)\,\mathbf{A_L}(t)$ is the quiver momentum imparted to the free electron by the laser field, and $\mathbf{A_L}(t)$ the potential vector associated to $\mathbf{E_L}$ by $\mathbf{E_L} = -\frac{1}{c}\frac{\partial}{\partial t}\mathbf{A_L}$

## 3. NUMERICAL RESULTS AND COMMENTS

Eq. (14) will be used to evaluate the transition probability of the ionization process under different conditions concerning the time pulse duration and the direction of the ejected electrons.

The case of temporal monochromatic radiation pulses has been addressed long ago [2], and will be briefly recalled in this communication. In the limit $t' \to -\infty$ and $t \to +\infty$, the transition amplitude given by Eq. (14), with the ground atomic state taken as $\left|\Phi_{i,0}(t)\right\rangle = e^{-iI_0 t}\left|\psi_0(\mathbf{r})\right\rangle$, assumes the form of a coherent sum of terms describing processes in which the atom is ionized by absorption of one x-ray photon together with exchange of n laser photons

$$A_{if} = \sum_{n=-\infty}^{\infty} \delta\left[\frac{k^2}{2} - \left(\omega_H + I_0 - U_p + n\omega_L\right)\right] \widetilde{A}_{i\mathbf{k}_n} \tag{16}$$

With $I_0$ the ground state energy, $U_p = \frac{E_{0L}^2}{4\omega_L^2}$ the ponderomotive shift,

$$\widetilde{A}_{i\mathbf{k}_n} = \frac{1}{2}\omega_L \int_0^{T_L} dt \exp\left\{ i\int^t \left[ \frac{\left[\mathbf{k}_n + \mathbf{k}_\mathbf{L}(\tau)\right]^2}{2} - \left(\omega_H + I_0\right) \right] d\tau \right\} \quad \left\langle \phi^-_{\mathbf{k}_n + \mathbf{k}_L}(\mathbf{r}) \middle| e^{-i\mathbf{k}_\mathbf{L}\cdot\mathbf{r}} \mathbf{E}_{0H} \cdot \mathbf{r} \middle| \psi_0(\mathbf{r}) \right\rangle \qquad (17),$$

evaluated at the canonical momentum $\mathbf{k}_L$, defined below, and $T_L$ the low-frequency field period.

By using Eq. (16) the differential probability per unit time of the ionization process with the exchange of n photons and ejection in the solid angle dΩ about the direction of the asymptotic drift electronic momentum $\mathbf{k}_n$ is easily derived as

$$\left(\frac{dP}{d\Omega}\right)_n = \frac{1}{2\pi} \left| \widetilde{A}_{i\mathbf{k}_n} \right|^2 k_n \tag{18}$$

with

$$\frac{k_n^2}{2} = \omega_H + n\omega_L + I_0 - U_p = \varepsilon_n \tag{19}$$

the drift kinetic energy of the ejected electron. The differential photoionization cross sections, obtained by dividing $\left(\frac{dP}{d\Omega}\right)_n$ by the incident x-ray photon flux, are shown in Fig. 1 for electron emission along the direction of the oscillating electric fields, for different values of the laser intensity. We note that, within the energy range shown in Fig. 1, for the laser parameters used in our calculations, the electron drift momentum $\mathbf{k}_n$ is greater than the quiver momentum amplitude.

The main contributions to the integral given by Eq. (17) come from the points of stationary phase satisfying dS/dt=0 with

$$S(t) = \frac{1}{2}\int^t \left\{ \left[\mathbf{k}_n + \mathbf{k}_L(\tau)\right]^2 - \omega_H - I_0 \right\} d\tau \tag{20}$$

implying, by assuming $\mathbf{E}_L = \hat{\mathbf{z}}\ E_{0L} \cos \omega_L t$,

$$\frac{1}{2}\left(\mathbf{k}_n + \mathbf{k}_{0L} \sin \omega_L t_s\right)^2 = \omega_H + I_0 \tag{20a}$$

with $\mathbf{k}_{0L} = -\frac{E_{0L}}{\omega_L}\hat{\mathbf{z}}$ and $t_s$ real. Therefore, for electron ejected along the laser electron field direction the highest values of the peaks occur at energies approximately given by $\frac{k^2}{2} = \frac{\left(\sqrt{2(\omega_H - I_0)} + k_{0L}\right)^2}{2}$ for the fastest electron, and $\frac{k^2}{2} = \frac{\left(\sqrt{2(\omega_H - I_0)} - k_{0L}\right)^2}{2}$ for the slowest electron. Through these values the spectrum width $\Delta\varepsilon$ can be estimated to be $\Delta\varepsilon \approx 2\sqrt{2(\omega_H - I_0)}k_{0L}$.

We observe that in both the reported spectra, for some values of the energy, some of the expected spikes characterizing the ejection probabilities are missing. This circumstance will be addressed below, after commenting on the spectra shown in Fig. 2 where the energy spectra for electron emission perpendicular to the direction of the radiation electric field are shown. The possibility of photoemission ejection at $\theta=\pi/2$, with $\theta$ denoting the angle between **k** and $\mathbf{E}_H$ , marks one of the difference with respect to the ordinary photoeffect. In fact, as it is well known, in the ordinary photoeffect ejection along the direction perpendicular to ionizing field direction is forbidden as the selection rules in dipole approximation require the angular momentum conservation and the constancy of the projection of the angular momentum along the field direction ($\Delta m$=0 with *m* the magnetic quantum number). For the case of our concern, starting from the ground state of hydrogen, only emission of electron into the partial Coulomb wave with orbital quantum number *l*=1 is permitted. Addition of the infrared field polarized along the direction of the x-ray allows the population of continuum states with $l\neq1$, due to infrared multiphoton exchanges, with the requirement of the constancy of *m* that, in our case, remain fixed to zero.

Because of this last restriction, due to the axial symmetry of the physical system, observation of electron with momentum directed along the perpendicular direction $\theta=\pi/2$ requires even parity of the orbital quantum number *l*. Therefore, together with the single x-ray photon absorption, only processes with exchange of an odd number of low-frequency photons enable the electron to be emitted in the direction perpendicular to that of the oscillating electric fields.

The energy spectra shown in Fig. 2 show sidebands evenly separated by $2\hbar\omega$. We remark that in our calculation we have ignored the contribution given by the field-induced atomic polarization. This correction could result to be comparable to the shown results. However, the qualitative aspects concerning photoemission perpendicular to the oscillating field direction continue to maintain their validity.

The effect of different XUV pulse durations is shown in Fig. 3. To obtain the results of Fig. 3 the amplitude probability given by Eq. (14) has been used, and the pulse electric fields have been assumed to have the following form

$$\mathbf{E}_j = \hat{\mathbf{z}}E_{0j}h_j(t)\cos^2\left[\pi\frac{t-t_{0j}}{\tau_j}\right]\cos\left(\omega_j t + \phi_j\right) \qquad (21)$$

where $h_j(t)=1$ for $t_{0j}-\tau_j/2<t< t_{0j}+\tau_j/2$ and zero elsewhere, $\tau_j$ is the total pulse duration, $t=t_{0j}$ corresponds to the peak of the pulse, $\omega_j$ and $\phi_j$ denote, respectively, the carrier frequency and the carrier-envelope phase of each pulse. Below the time difference $t_{0L}$-$t_{0H}$ between the peaks of the envelopes will be denoted by $\Delta T_H$. Moreover, in our calculations, $t_{0L}=0$, $\phi_H=0$ and $\phi_L=\pi$.

In order to have an integer number of cycles, the pulse duration has been taken as $\tau_{H,L}=N_{H,L}T_{H,L}$ with $T_{H,L}=2\pi/\omega_{H,L}$. By choosing high values of $N_L$ ($N_L\approx500$) the low-frequency pulse may be considered a monochromatic field.
Of course, by decreasing the pulse duration the peaks broadening increases according to the energy-time uncertainty relation $\Delta E\Delta t\sim\hbar$. In Fig. 3a, the arrows on the energy axis mark the energy values in correspondence of very small values of the ejection probability. We note that for the cases shown in the figures, when the pulse duration varies, the positions of the energy corresponding to the lack of the peaks keep fixed, being the same as the ones found when the ionization is produced by the action of monochromatic radiation fields. This circumstance, as well as the whole structure of the energy spectra, may be better understood by describing the photoemission event in terms of interferences of the transition amplitude in the time domain. This kind of approach has been followed in ref. [13] and [14] for explaining, respectively, the presence of structures in the photoemission energy spectra measured in the photodetachment experiments of $F^-$ carried out with linearly [15] and circularly [16] polarized radiation field. To make more evident the role played by the interference in the time domain, let us consider the transition probability by assuming a rectangular x-ray pulse, beginning at t=0, whose duration $\tau_H=\nu T_L$ encompasses $\nu$ low-frequency laser cycles. Dividing the time interval into the $\nu$ cycles, the ionization probability at time $\tau_H$ due to the absorption of one x-ray photon, by putting in Eq. (14) t'=0, takes the form

$$\frac{dP}{d\varepsilon_k d\Omega}=\left|A_{i\mathbf{k}}(1)\right|^2\sqrt{2\varepsilon_k}\,\frac{\sin^2\nu\pi\Delta}{\sin^2\pi\Delta} \tag{22}$$

Where $A_{i\mathbf{k}}(1)$ is the transition amplitude given by Eq. (17) evaluated at the canonical momentum **k**, $\varepsilon_k=k^2/2$ takes continuous values, and

$$\Delta=\frac{1}{\omega}\left[\frac{k^2}{2}-\left(\omega_H+I_0-U_p\right)\right]. \tag{23}$$

According to Eq. (22) the probability $dP/(d\varepsilon_k d\Omega)$ shows an N-slit interference pattern with the principal maxima, obtained taking $\pi\Delta=n\pi$, located at energies $\varepsilon_n$ given by Eq. (19). The height of the principal peaks is modulated by the factor $|A_{i\mathbf{k}}(1)|^2$ that plays the role of a diffraction function. Therefore, the peak suppression occurs at energy values where the function $|A_{i\mathbf{k}}(1)|^2$ becomes vanishingly small. The role played by the duration of the XUV pulse is illustrated in Fig. 3b where the energy spectra calculated by means of Eq. (14), with the electric fields given by Eq. (21), are shown for $\tau_H=10T_L$ and $\tau_H=3T_L$.

In the presence of a monochromatic low-frequency radiation pulse, the spectra turn out to be essentially equal to that already shown in Fig. 3a. By shortening the duration of the low-frequency field pulse, the shape of the extreme wings of the energy spectra turns out to be not altered, while in the intermediate energy region the spectra are smeared out, as shown in Fig. 3b. Calculations here not reported show that when the XUV pulse encompasses several low-frequency periods ($\tau_H>=3T_L$) the energy spectra do not change appreciably provided the time lag $\Delta T_H$ between the two radiation pulses is less than or equal to $T_L/4$. Instead, when the XUV pulse duration is twice the low-frequency period, notable changes appear in the photoelectron energy spectra calculated respectively at $\Delta T_H=0$ and $\Delta T_H=T_L/4$ (see fig.4). For pulse duration equal to the low frequency field period the differences becomes very marked, as shown in Fig. 5.

In order to get more insight into this last feature we use the stationary phase method for calculating the amplitude probability $A_{i\mathbf{k}}$ of Eq. (14). It turns out that $A_{i\mathbf{k}}$ may be approximated by the sum of the two contribution evaluated at the couple of instant $t_1$ and $t_2$ determined, for assigned k, by Eq. (20a)

$$A_{i\mathbf{k}} \approx i\sqrt{\frac{2\pi}{\ddot{S}}}e^{i\frac{\pi}{4}}M(t_1)E_{0H}\left[f_H(t_1)e^{iS(t_1)} + f_H(t_2)e^{i\left(S(t_2)-\frac{\pi}{2}\right)}\right] \tag{24}$$

$f(t) = h_H(t)\cos^2\left[\pi\frac{t+\Delta T_H}{\tau_H}\right]$, $M(t) = \left\langle \phi^-_{\mathbf{k}+\mathbf{k}_L}\middle| e^{i\mathbf{k}_L(t_1)\cdot\mathbf{r}} z \middle|\psi_0\right\rangle$, $|\ddot{S}| = |\ddot{S}(t_1)| = |\ddot{S}(t_2)|$. As $\omega_L(t_1+t_2)=-\pi$, it is very easy to show that $M(t_1)=M(t_2)$. We note that for $t_1=-\pi/(2\omega_L)$, $\ddot{S}=0$ and the approximation given by Eq. (24) is no longer valid.

By Eq. (24), with the above limitation, the differential transition probability may be written as an oscillating function

$$\frac{dP}{d\varepsilon_k d\Omega} = \frac{2\pi}{|\ddot{S}|}|M|^2\left[f_H^2(t_1) + f_H^2(t_2) + 2f_H(t_1)f_H(t_2)\sin|S(t_1) - S(t_2)|\right]k \tag{25}$$

Comparison in Fig. 6 between Eq. (18) and Eq. (25) obtained, respectively, by numerical evaluation of the integral entering Eq. (17) shows that the representation of $A_{if}$ given by Eq. (24) fails at the values of the photoelectron energies in correspondence of the absolute maxima of the differential transition probability. Nevertheless, Eq. (25) catches the oscillating behavior of $(dP/d\varepsilon_k d\Omega)$ as function of the energy of the ejected electron. By increasing the time lag from $T_L=0$ to $T_L=\pi/4$ the oscillating amplitudes increase, and for $\Delta T_H=T_L/4$, being $f_H(t_1)=f_H(t_2)$, the differential probability amplitude vanishes for $\sin[S(t_1)-S(t_2)]=-1$. The inset in Fig. 6 shows the couple of instants, calculated by Eq. (20a), in correspondence of a zero of $(dP/d\varepsilon_k d\Omega)$. For different time lags, $f_H(t_1)\neq f_H(t_2)$ and $(dP/d\varepsilon_k d\Omega)$ oscillates, due to the presence of the interference terms in Eq. (25), without vanishing.

The case of ionization in the presence of an infrared radiation by an x-ray pulse, whose duration has been taken of the order of a fraction of the IR radiation period, has been subject of both theoretical and experimental investigations [10,11]. It has been proposed to get information on the temporal behavior of the low-frequency field.

For pulse duration less than, or of the order of, $T_L/4$ the single peak of the electron energy spectrum is located at a value of the energy that depends on the amplitude of the vector potential associated to the low-frequency field at the instant of the ionization event, when the intensity of the ionizing XUV pulse reaches its maximum value. By denoting with $\boldsymbol{\pi}_0$ the field-free mechanical momentum of the electron freed by the sole absorption of the x-ray photon, in absence of the low-frequency field, it turns out that when the low frequency radiation field is switched on, due to the conservation of the electron canonical momentum, under the electric dipole approximation, the electron energy at the end of the low frequency pulse is given by $\varepsilon_k=[\boldsymbol{\pi}_0+\mathbf{k}_L(\tau_0)]^2/2$, where $\tau_0$ is the instant at which the XUV has its maximum intensity.

By carrying out successive measurements with different delays between XUV and low frequency fields it is possible to get information on $\mathbf{k}_L(\tau)$ and, then, on its numerical derivative with respect to the time, giving $\mathbf{E}_L(t)$. In fact, upon dividing electron energies differences measured in two successive instants by the elapsed time, information about the temporal behavior of the low-frequency laser field may be obtained

$$\frac{1}{\delta t}\left[\varepsilon_k(t+\delta t/2)-\varepsilon_k(t-\delta t/2)\right]\approx \boldsymbol{\pi}_0\cdot\mathbf{E}_L(\tau) \qquad (26)$$

In the present paper we outline an alternative method that do not require knowing $\boldsymbol{\pi}_0$ for the determination of $\mathbf{E}_L(t)$. In fact, through simultaneous energy measurements of the electrons ejected in opposite directions, along the fields, by using

$$\sqrt{2\varepsilon_{k\pm}(\tau)}=\pi_0\pm k_L(\tau) \qquad (27)$$

the instantaneous values of the quiver electron momentum may be obtained as

$$k_L(\tau)=\frac{1}{2}\left(\sqrt{2\varepsilon_{k+}(\tau)}-\sqrt{2\varepsilon_{k-}(\tau)}\right) \qquad (28)$$

Fig. 7 shows two electron energy spectra obtained by numerical integration of the tridimensional time-dependent Schroedinger equation describing the ionization process of our concern when the duration of the XUV pulse, embracing 20 HF radiation cycles, is taken equal to $T_L/3$. The integration has been carried out by using the QPROP program of ref. [17].

Comparison with numerical calculations based on Eq. (14) shows very good agreement with the theoretical approach outlined in section 2. It is evident how the difference between the energy values located at the peaks of the energy spectra varies by varying the delay between the fields. By successive measurements of these differences for different time lags, the oscillating electron momentum as a function of time may be obtained according to Eq. (28).

By concluding, we have studied the two color ionization process of hydrogen by solving approximately the Schroedinger equation. Electron energy spectra have been calculated for three different regions of the x-ray pulse duration. When the x-ray pulse encompasses several low-frequency radiation cycles, the energy spectra exhibit peaks that may be explained in terms of quantum interference in the time domain. Analogously to the optical n-slit diffraction the peaks intensity is governed by a diffraction function whose zeros determine the energies at which the ionization probabilities become vanishingly small. For x-ray pulse duration of the order of a LF field period, the photoelectron energy spectra shows a completely different behavior. Oscillations appear whose amplitude is defined by the time lag between the x-ray pulse and the low-frequency field. These findings could be confirmed by experiments similar to those reported in refs. [6,11] carried out with XUV pulses encompassing a single laser cycle, while, by using many laser cycles XUV pulses, lack of some peaks in the photoelectron energy spectra, as shown in Fig. 3, should be observed.

Finally, for very short x-ray pulse containing however several cycles, the ionization event may be thought to occur instantaneously. The energy location at which the ionization probability reaches a peak depends on the "instantaneous" value of the quiver momentum imparted to the electron by the low-frequency field. This circumstance, already exploited to retrieve the temporal shape of the assisting laser field, suggests an alternative way of determining the shape of the oscillating laser field, by simultaneous measurement of the energy spectra of the electron ejected in opposite direction.

.

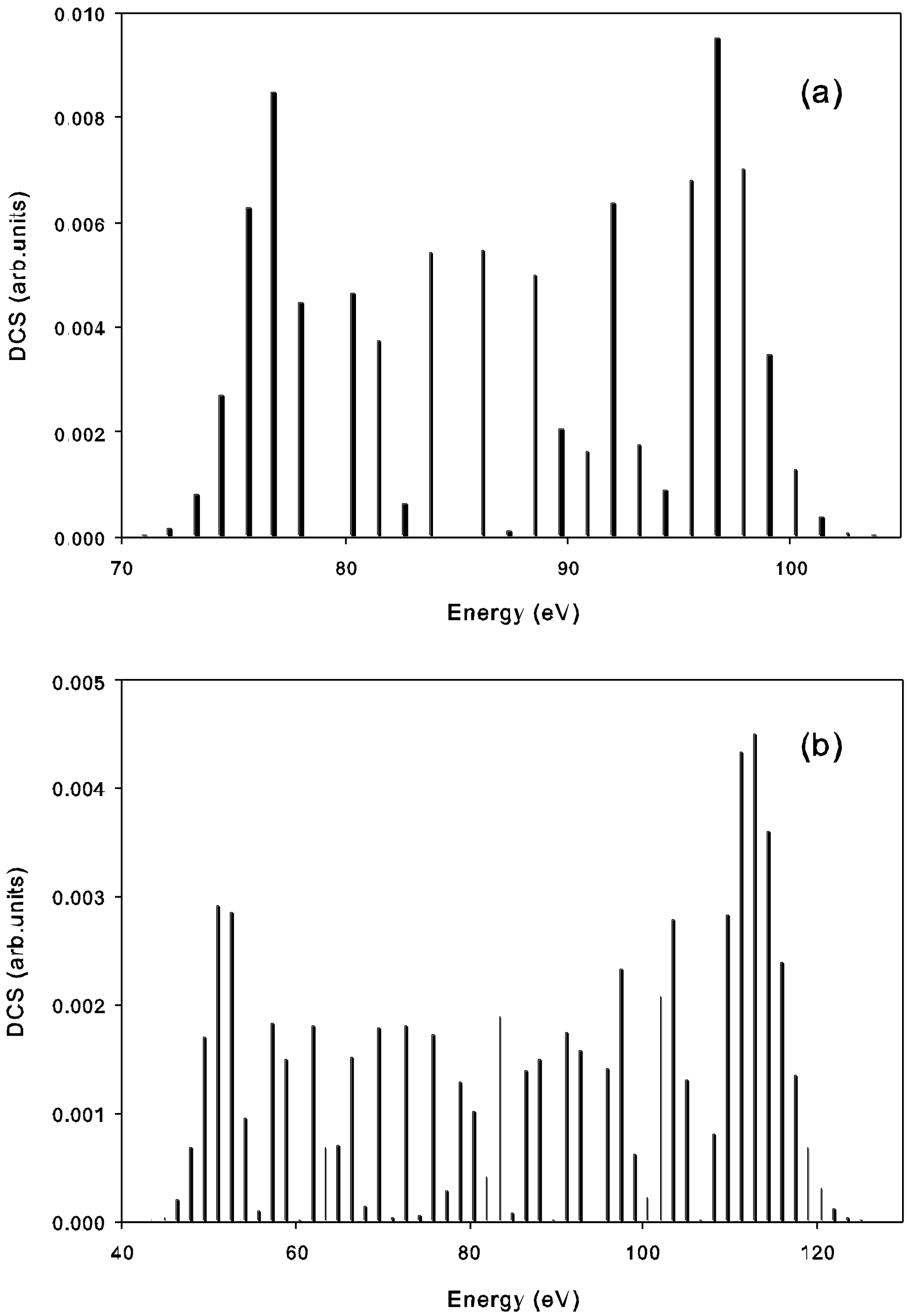

Fig. 1 Photoionization differential cross section (DCS) in arbitrary units *vs* the photoelectron energy for two different values of the XUV and low-frequency photon energies and laser intensity.

(a) laser intensity $2*10^{12}$ W/cm$^2$ ; low-frequency photon energy $\hbar\omega$=1.17 eV; XUV photon energy $\hbar\omega$=100 eV.

(b) laser intensity $3*10^{13}$ W/cm$^2$ ; low-frequency photon energy $\hbar\omega$=1.55 eV; XUV photon energy $\hbar\omega$=93 eV. For both cases the electron ejection direction is parallel to the oscillating fields ($\theta$=0).

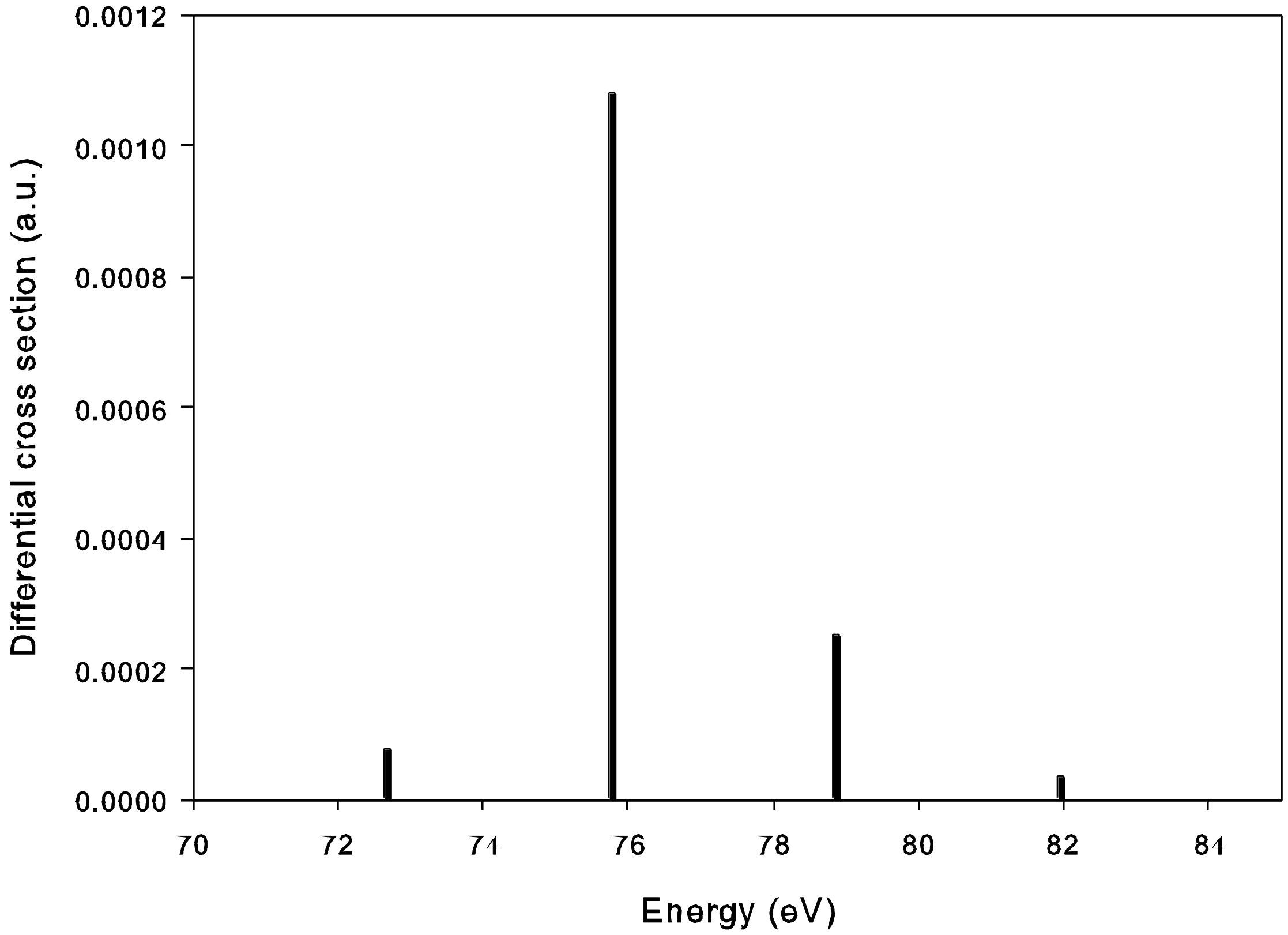

Fig. 2 Photoionization differential cross section vs the photoelectron energy.
Laser intensity $I=3*10^{13}$ W/cm$^2$; low-frequency photon energy $\hbar\omega=1.55$ eV. Ejection angle $\theta=90$ degrees. The peaks are spaced by 2 $\hbar\omega_L$

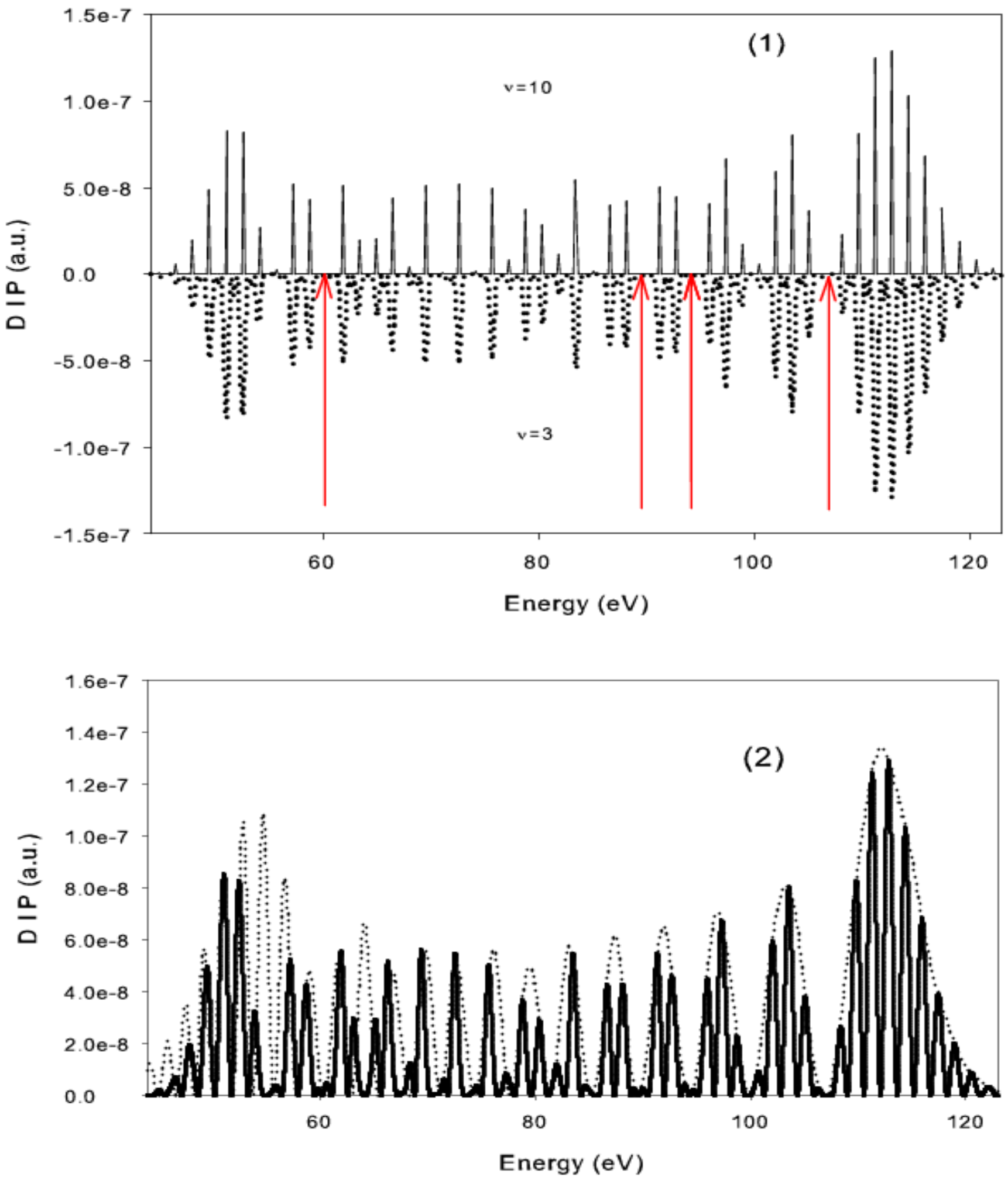

Fig3a

1) Differential Ionization Probability (DIP), in atomic units, divided by $\nu^2$ as a function of the photoelectron energy. $\nu$ is the number of the laser field periods encompassed in the XUV pulse. XUV intensity: $10^{11}$ W/cm$^2$; Laser intensity: $3{\cdot}10^{13}$ W/cm$^2$; XUV photon energy: 93eV; laser photon energy: 1.55 eV. The angle between the electron asymptotic momentum and the oscillating electric filed is $\theta$=0 degrees. The arrows marks the energy values in correspondence to vanishing ionization probability. A rectangular XUV pulse shape has been assumed. The LF pulse may be considered to be monochromatic ($\tau_L$ = 500$T_L$)

2) As in 1). Full line: $\nu$=2; dotted line: diffraction function $|A_{1f}|^2$ evaluated with $\nu$=1. Note that the diffraction function is the envelope of the spectrum and that for some photoelectron energy the zeros of $|A_{1f}|^2$ fall on the expected peaks location.

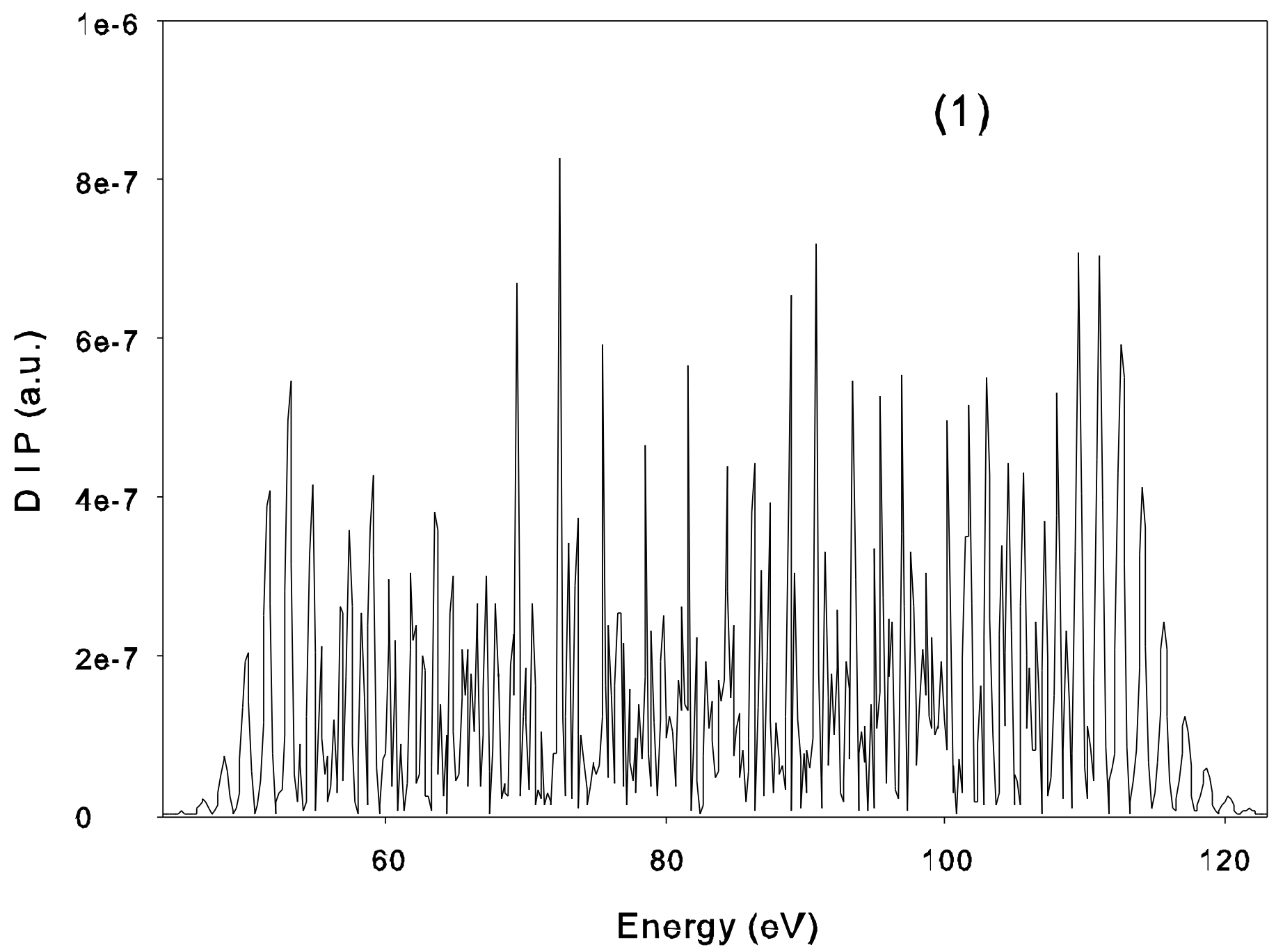

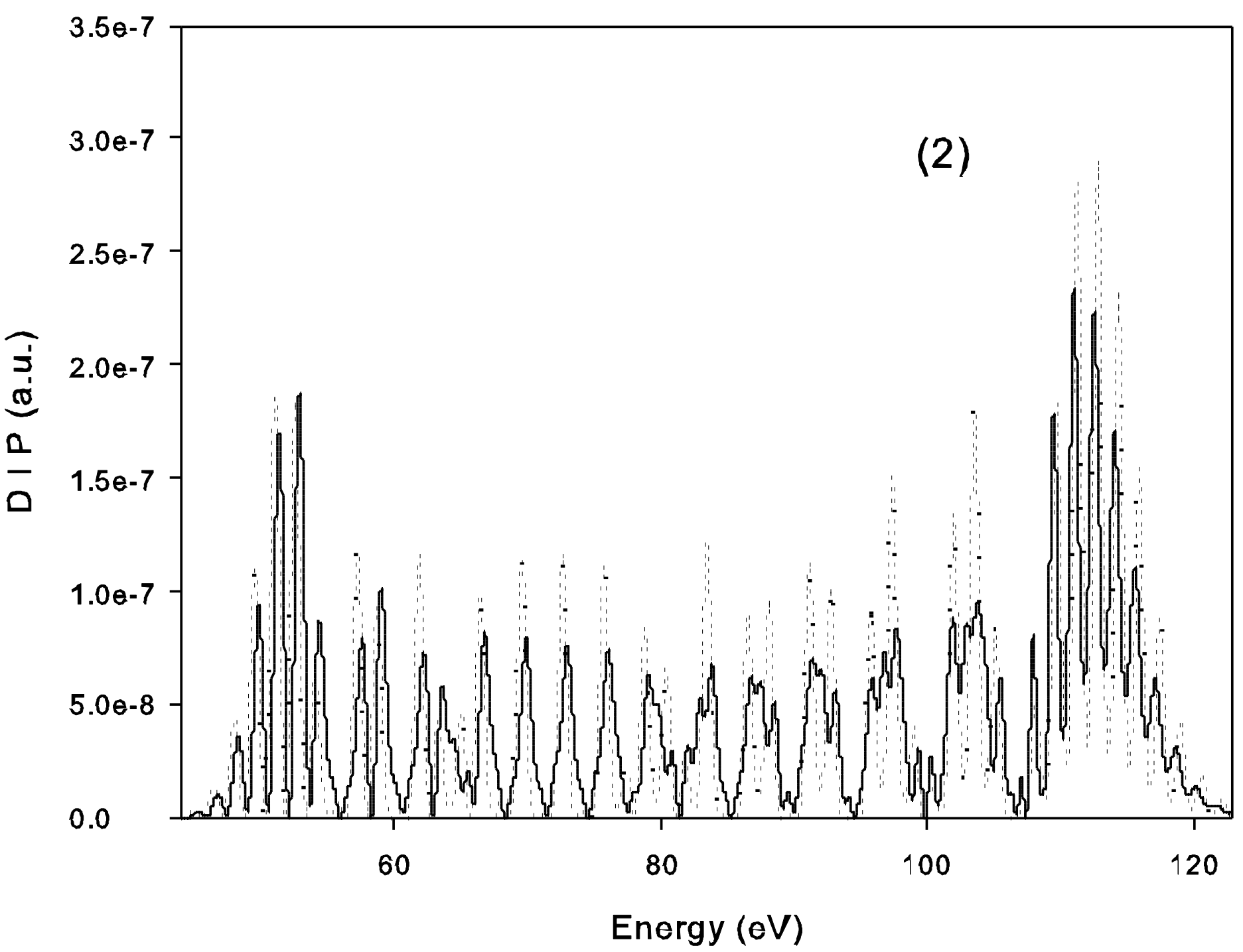

Fig 3b

1) As in Fig 3a with $\tau_L$ and $\tau_H$ equal to $10T_L$.

2) As in Fig. 3a with $\tau_L=10T_L$ and $\tau_H=3T_L$. The dotted lines, shown for comparison, are the DIP of Fig.3a1 with $\nu=3$.

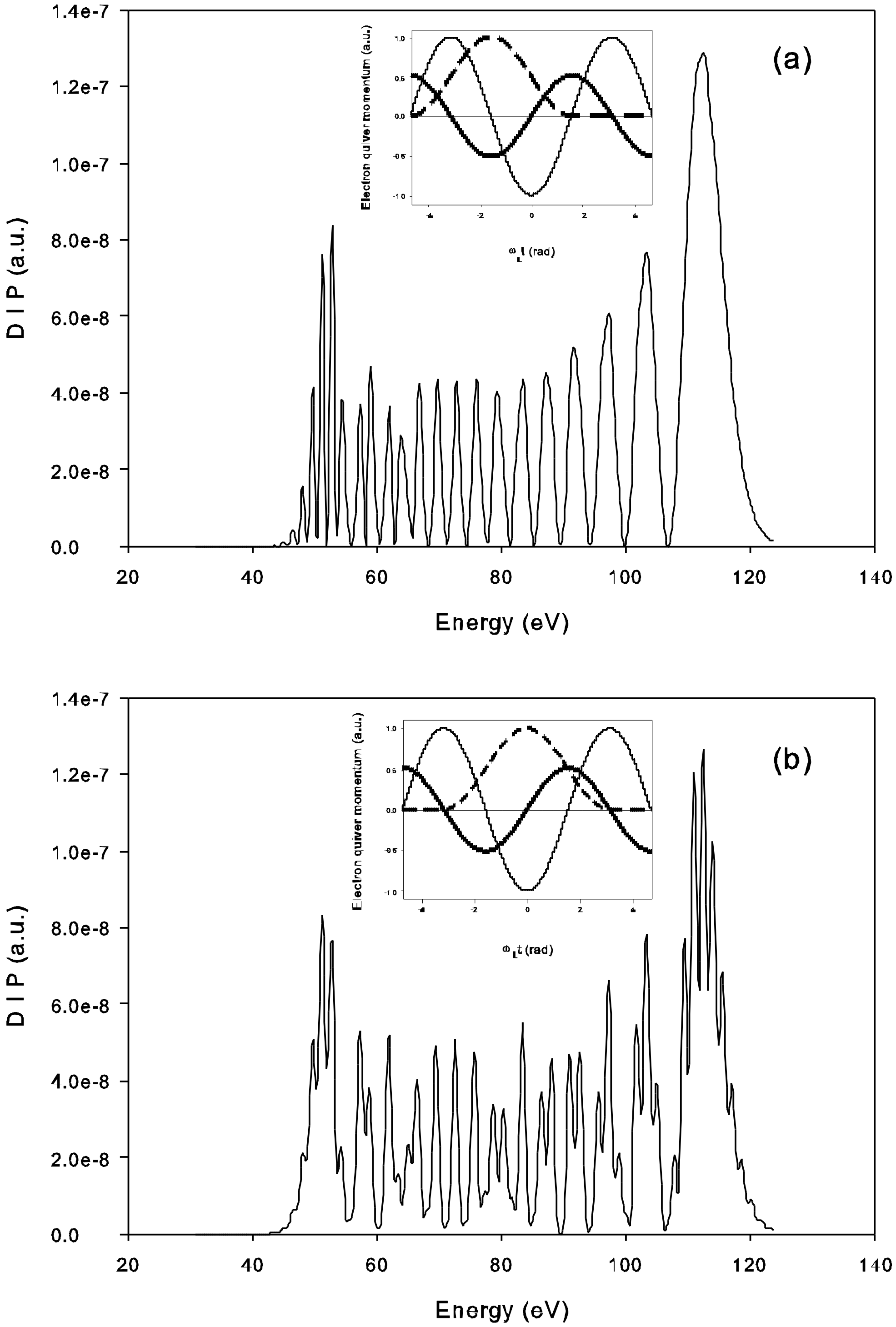

Fig.4. As in Fig. 3a with $\nu=2$. (a) Time lag $\Delta T_H = T_L/4$; (b) time lag $\Delta T_H=0$. The insets show the XUV pulse envelope (dashed line), the quiver electron momentum in atomic units (thick line) and the low frequency electric field in arb. units (thin line) as a function of time.

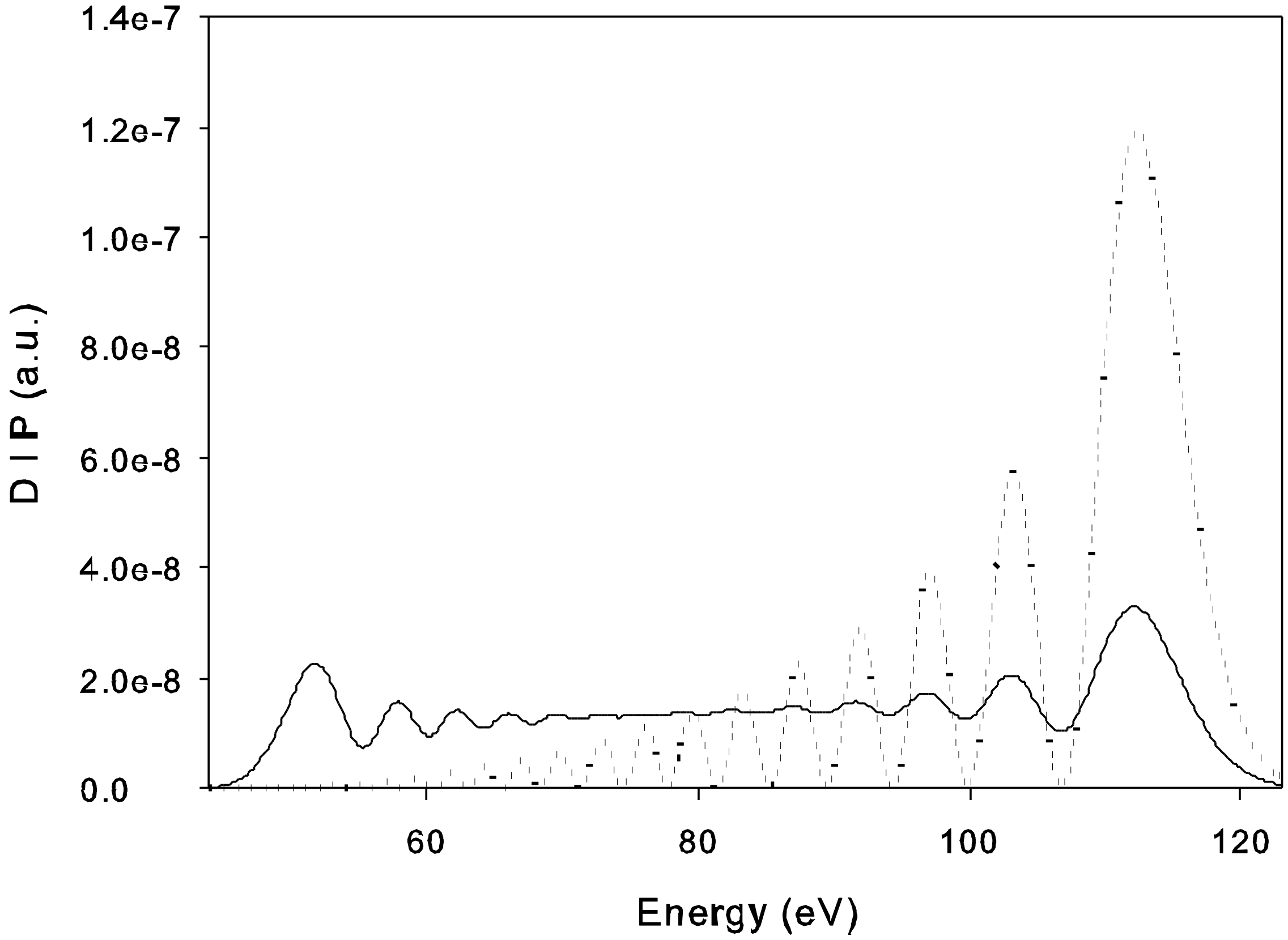

Fig.5 As in Fig.3a, for two different time lags and XUV pulse duration equal to the laser period.

Dotted line: time lag $\Delta T_H = T_L/4$; Full line: $\Delta T_H = 0$.

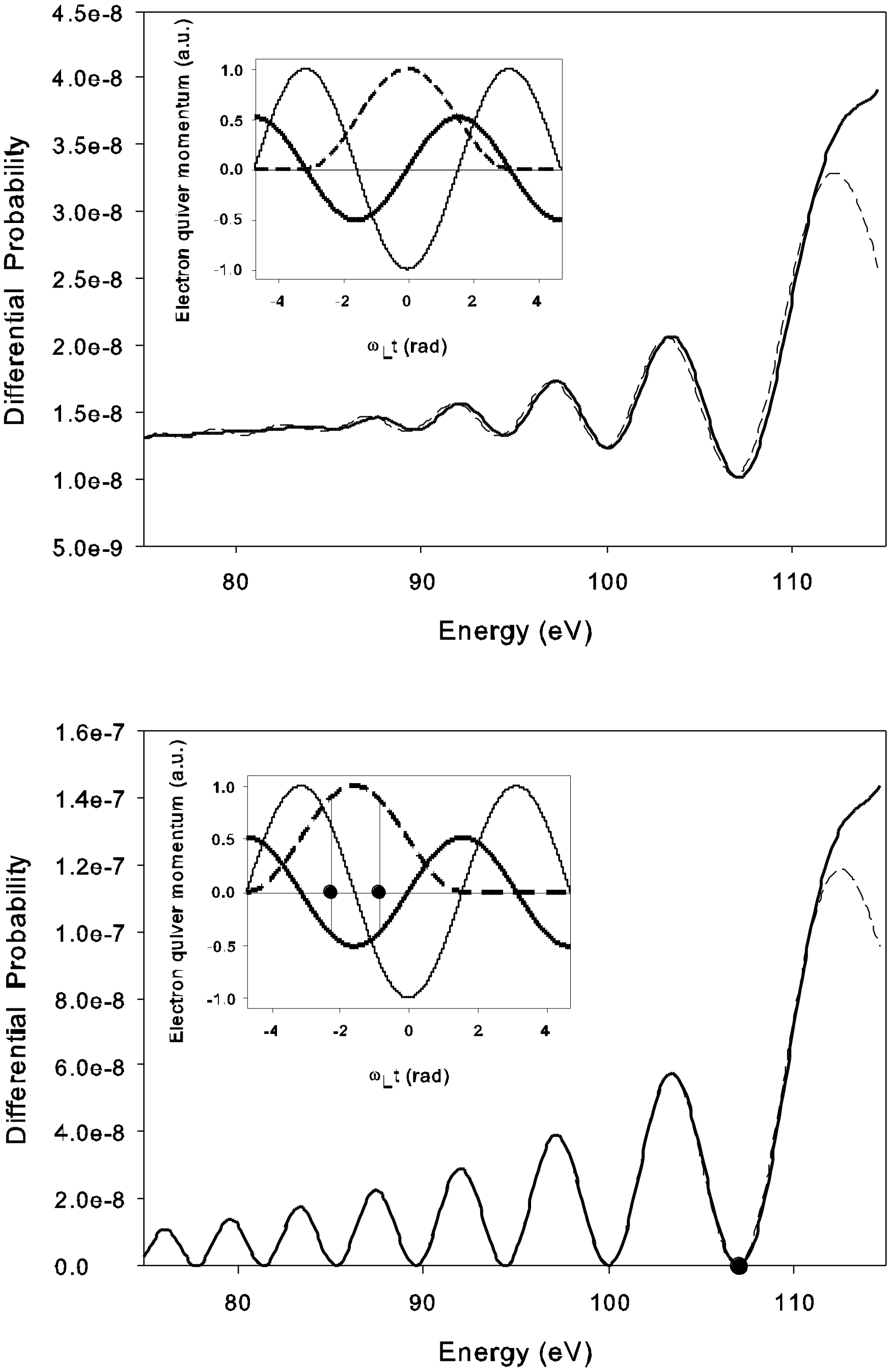

Fig.6

a) Comparison of the photoemission differential probability for $\Delta T_H=0$ as a function of the electron energy between the results based, respectively, on numerical evaluation of the integral of Eq. (14) (dashed line) and on the stationary phase method (full line). Note the asymmetrical disposal of the XUV envelope with respect to the curve showing the electron quiver momentum.

b) As in a) for $\Delta T_H=T_L/4$. At the stationary points marked in the inset, $\sin[S(t_1)-S(t_2)]=-1$ (See eq.(25)), and the DIP vanishes at the energy of 107 eV marked on the energy axis. Laser and XUV parameters as in Fig.5

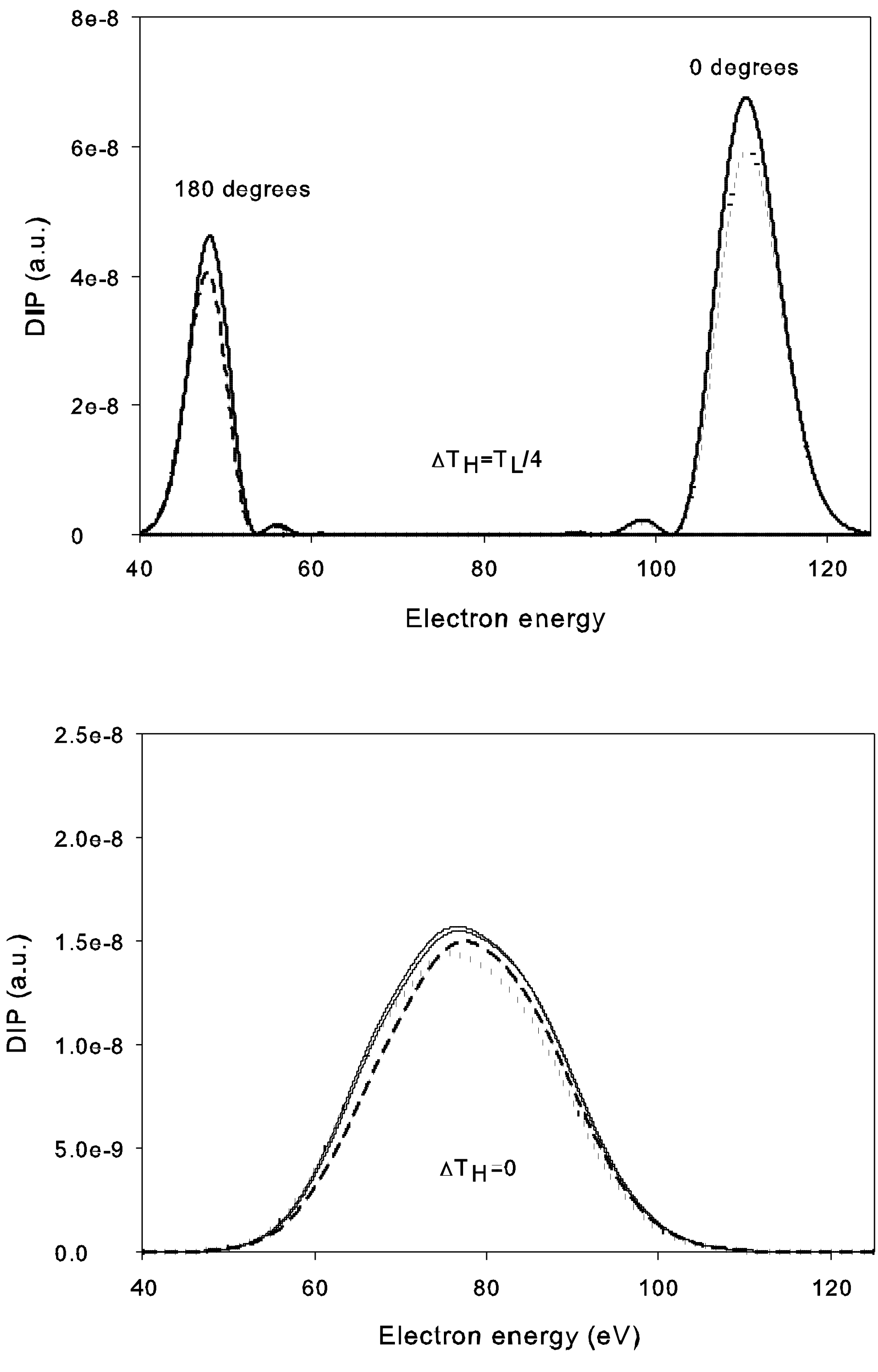

Fig.7. Differential ionization probability, for opposite electron ejection directions (0° and 180°), evaluated at different time lags. Laser and XUV Intensities and frequencies as in Fig.3a. Shown are too the results of the tridimensional numerical integration of the Schroedinger equation (TDSE). Full lines Eq. 18, dashed line TDSE $\theta$=180°, dotted line TDSE $\theta$=0°. The time lags are indicated on each panel. The XUV pulse duration is $T_L/3$